# Modulating the electronic and optical properties of a ternary chalcogenide CdTl$_2$Te$_4$ via external electric field: A DFT study

Aditya Dey

*Department of Physics, Indian Institute of Technology Patna, Bihta Campus, Patna, Bihar, India-801106*

Various ternary chalcogenide systems and their properties are one of the hot topics for researchers nowadays. In this article, one of the ternary chalcogenide compounds, CdTl$_2$Te$_4$ is studied including its electronic structures and optical properties via first principles calculations using SIESTA code. The structure of the compound has a tetragonal crystal system and the unit cell is periodic in all directions. The band structure showed it is a direct band gap semiconductor which is explained by the density of states. The energy gap obtained is seen to increase on employing external electric field perpendicular to the xy plane, but does not increase on increasing the magnitude of field strength. The optical properties which includes absorption coefficient, reflectance, real & imaginary parts of dielectric function, refractive index and extinction coefficient were calculated for polarized light both in plane ($\mathbf{E}\perp c$) and out of the plane ($\mathbf{E}\parallel c$), but the interesting modulation in properties were seen for the latter which can open up applications of the compound in UV absorbers, making solar cells, etc. So, the compound being a narrow-gap semiconductor along with its tunable electronic and optical properties can have potential application in the fields such as optoelectronics.

**Introduction**

With every passing day, researchers around the globe are carrying out extensive study on various material elements and compounds and trying to investigate their properties so that they can be of use in different fields. As we all know, from the past decade, two-dimensional materials are one of the most studied materials due to their applications in various fields. 2D materials like graphene, phosphorene, silicone, stanene, germanene, etc have properties that can be tuned and are being used in numerous fields [1-4]. Also there are transition metal dichalcogenides which have 2D layered structure, like MoS$_2$, WSe$_2$, TiTe$_2$ and many more [5-7] showing remarkable properties. Apart from these materials, there are other materials and compounds that can be competent enough to the 2D materials. Of these, ternary and quaternary chalcogenides have attracted researchers due to their structure types and properties. Although, study on these compounds had started way back [8-11], modern day researchers are studying them more meticulously and also have made recent advances [12-14]. AB$_2$X$_4$ type (A & B are metals and X is a chalcogen atom) ternary chalcogenides (TCs) is one of the kinds that have redundant physical properties and various applications [15-18]. There are several types of ternary chalcogenide compounds of this type that are previously studied theoretically and characterized and synthesized experimentally as well [19-20]. The metal chalcogenides are quite different from the metal oxides and show incredible behaviours mainly because of their tendency to crystallize in low-dimensional layered structures [21]. These low-dimensional and anisotropic properties gives rise to various physical properties like superconductivity and charge density wave formation, which is also seen in $AB_2X_4$ type TCs, as reported by many researchers [22-24].

In chalcogenides, the repulsions between the chalcogen atoms are small and the structures are stabilized by the van der Waals attraction of these chalcogen atoms, which is more predominant in the tellurides [21]. Also, Te atom having larger ionic radius than S and Se, the tellurides have a rich structural chemistry including different electronic and physical properties. So, one of the tellurides, $CdTl_2Te_4$ which is a $AB_2X_4$ type TC, is studied in this article. This compound has been studied previously almost four decades ago by S.K Karimov which included study of various properties and applications of the compound [25-31]. The findings show that the compound has tunable properties like solubility, hardness, thermoelectric and thermodynamic properties, electrophysical properties, etc at finite temperature which is achieved by various methods like doping. Charge carrier recombination and dielectric permeability of $CdTl_2Te_4$ were also studied [32]. With such interesting properties of the compound and the recent findings of researchers on this type of TCs, the electronic and optical properties of $CdTl_2Te_4$ is investigated by ab-initio density functional theory calculations and reported in this article. Further, in order to modulate the optical properties and tune the band structure, external electric field is also applied.

**Computational methods**

The calculations are done using ab-initio density functional theory (DFT) as implemented in the software SIESTA [33-37]. A real space mesh cutoff of 300 Ry has been used and the generalized-gradient approximation (GGA) with the Perdew-Burke-Ernzerhof (PBE) form is chosen for the exchange-correlation functional [38]. Normconserving pseudopotentials in the fully nonlocal Kleinman− Bylander form have been considered for all the atoms [39]. Firstly, the unit cell of the compound is taken and optimization is done followed by calculating the electronic structure and optical properties. For the optimization and calculation of these properties, a 15 X 15 X 15 k-point grid has been considered within the Monkhorst−Pack scheme [40]. Further, to study the effect of electric field along out of the plane direction, a vacuum of 18 Å is created in the unit cell along z direction or *c* axis and the properties were calculated using self-consistent calculations. For this, the k-point grid considered is 20 X 20 X 1. The convergence for energy was chosen as $10^{-4}$ eV between two steps. The optical properties properties were calculated for in plane ($\mathbf{E} \perp c$) and out of plane ($\mathbf{E} \parallel c$) polarized light for an energy range of 0-30 eV, using an optical broadening of 0.1 eV. The optical mesh taken is 20 X 20 X 20 and 40 X 40 X 1 for optical properties without and with electric field. Conjugate-gradient (CG) method [41] has been used to optimize the structures and a double-ζ polarized (DZP) basis set is used. Systems are considered to be relaxed only when the forces acting on all the atoms are less than 0.01 eV Å$^{-1}$.

**Results and discussions**

The $AB_2X_4$ type TC compound, $CdTl_2Te_4$ crystallizes in the tetragonal crystal system with a space group $I\bar{4}2m$, stable at atmospheric pressure. The primitive or unit cell of the compound which is periodic in all directions is considered firstly. Figure 1 shows the optimized crystal structure that includes the symmetrized primitive cell and the periodicity in all directions is illustrated by drawing an additional unit in each of the directions followed by the formation of 2x2x2 supercell of the compound. The structure shown is based upon a cubic close packed array of Te atoms with the Tl atoms occupying a fraction of available tetrahedral sites and Cd atoms occupying the corner position of the lattice. The unit cell has 7 atoms including two Tl atoms, one Cd and four Te atoms. The lattice cell parameters and cell angles of the unit cell are $a = 6.565$ Å, $b = 6.565$ Å, $c = 8.005$ Å, $\alpha = 113.893°$, $\beta = 113.886°$, $\gamma = 90.024°$. The coordinates of the atoms are mentioned in Table 1. The bonds formed are Tl-Te and Cd-Te and their respective bond lengths are 2.92 Å and 2.88 Å. Further, for studying the effect of electric field, the structure is created such that it is periodic along x and y-directions ($a$ and $b$ axis) and vacuum along $c$ axis. The external electric field ($E_z$) is then applied along z-direction, perpendicular to the xy plane.

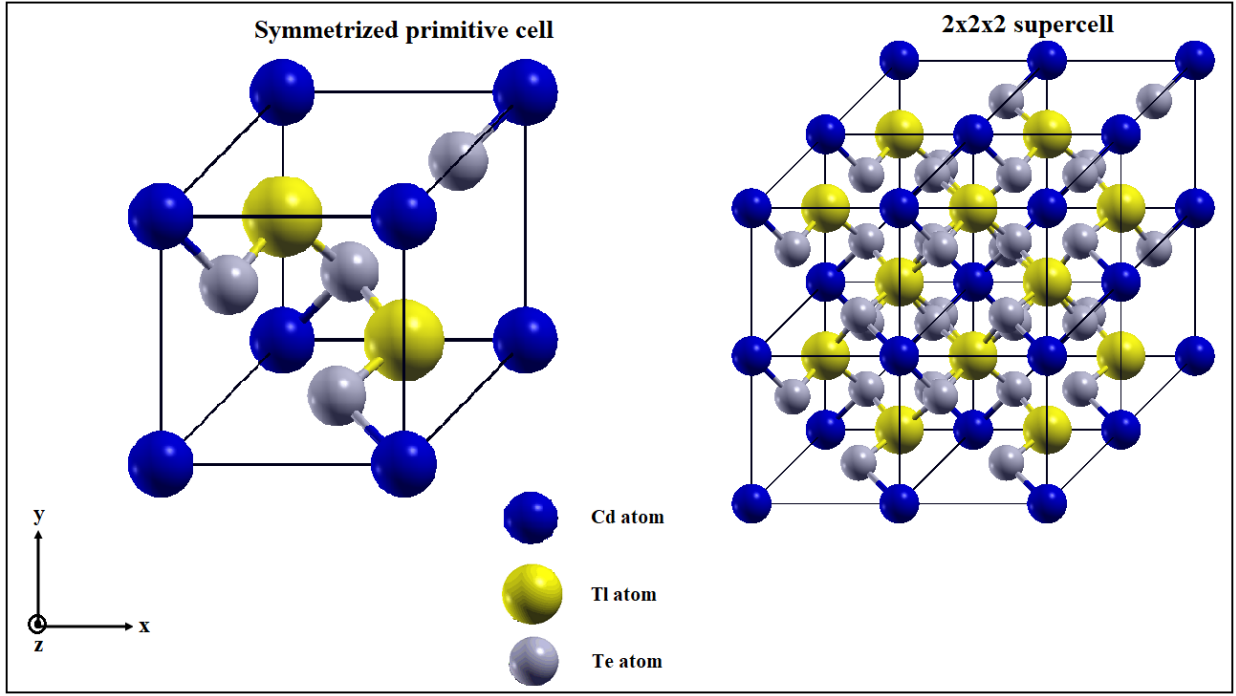

Figure 1: Optimized structures of symmetrized primitive cell and 2x2x2 supercell of CdTl$_2$Te$_4$

| Atoms | x(Å) | y(Å) | z(Å) |
|---|---|---|---|
| Tl | 0.000 | 3.209 | 3.240 |
| Tl | 3.222 | 0.000 | 3.232 |
| Cd | 0.000 | 0.003 | 0.001 |
| Te | 1.761 | 1.795 | 1.386 |
| Te | 4.639 | 4.653 | 1.394 |
| Te | -1.412 | 1.430 | 5.102 |
| Te | 1.401 | -1.448 | 5.099 |

Table -1 Atomic coordinates of the unit cell of CdTl$_2$Te$_4$ in Å

**Electronic structures**

Taking the optimized geometry of the unit cell, the electronic structures of the compound are calculated. The band structure and density of states (DOS) of CdTl$_2$Te$_4$ is shown in Figure 2. The brilloiun zone taken is of the body centered tetragonal lattice. It can be seen that the compound behaves as direct gap semiconductor with a narrow band gap ($\Delta_g$) of 0.32 eV. The conduction band bottom (CBB) and valence band top (VBT) are found at the $\Gamma$ point only every time it is considered in the path of high symmetry points taken (Figure 2). The DOS plot in Figure 2 shows there are no states present near the Fermi level suggesting an energy gap between the conduction and valence bands. To understand the electronic structures better, partial density of states is also plotted (PDOS) shown in Figure 3. The electronic configuration for Cd is [Kr] $4d^{10}5s^2$, Tl is [Xe] $4f^{14}5d^{10}6s^26p^1$ and Te is [Kr] $4d^{10}5s^25p^4$. So the PDOS is plotted for 4d and 5s orbitals in case of Cd, 6s and 6p for Tl and 5s and 5p for Te. The plot shows that the VBT is mainly due to Te 5p and Tl 6p states and minor contribution from Cd 5s and 4d states. The CBB is predominantly due to Te 5p and Tl 6s states with some contribution from Tl 6p states. The states in the valence band is mainly due to Te 5p states and some due to Tl 6p states from the Fermi level itself with Cd 5s contributing at -3.5 eV and Tl 6s at -6 eV. There is a hint of hybridization of Tl 6s and Te 5p states in both valence and conduction band. In conduction band, Tl 6p along with Te 5p majorly is responsible for the states in conduction band with Cd 5s also at 3 eV. It can be seen Te 5p and Tl 6p states play an important for the band gap here.

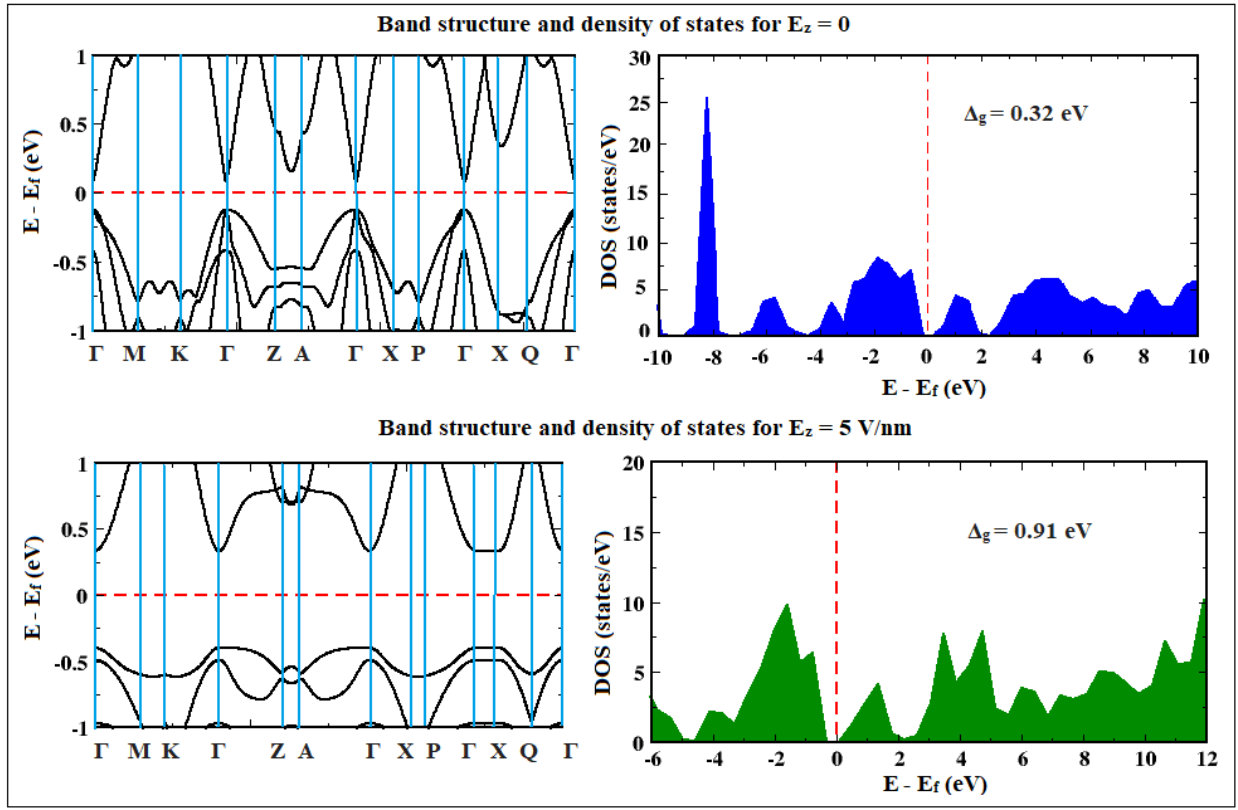

Figure 2: Band structures and density of states without $E_z$ and with $E_z$ = 5 V/nm. Plots are scaled with Fermi energy ($E_f$) and red dotted line shows the Fermi level

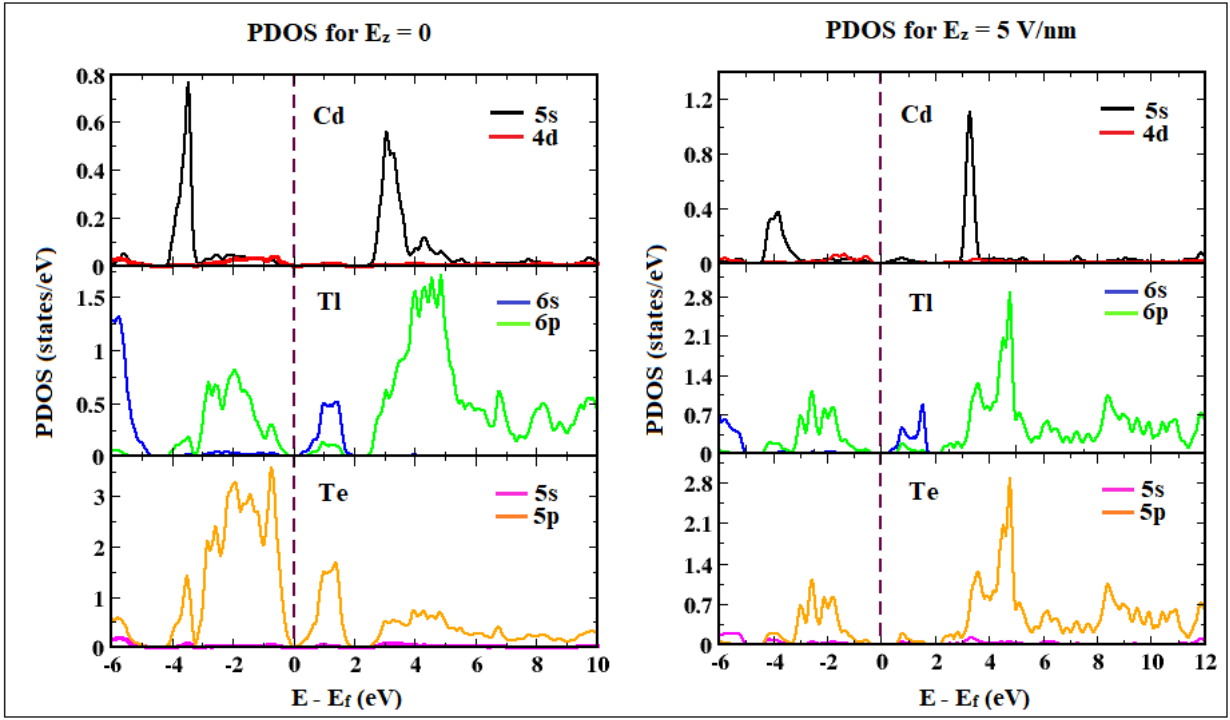

Figure 3: Partial density of states (PDOS) for E = 0 and E = 5 V/nm. Plots are scaled with Fermi energy ($E_f$) and dotted line shows the Fermi level

Next, when $E_z$ is applied, an increase in the band gap can be seen. However, the increase in $\Delta_g$ is sudden and it does not alter much with increase in field strength. From Figure 4 it can be observed that $E_z$ of strength 1-10 V/nm is applied and the $\Delta_g$ increases from 0.32 to 0.92 eV at 1V/nm and remains almost same. The band structure and DOS for $E_z$ = 5 V/nm is shown in Figure 2. It can be noticed that both the CBB and VBT move away from the Fermi level thus enabling a gap to open. This is seen because the electric field applied tends to trigger the charge density of the atoms which also increases the orbital excitation of electrons resulting in the increase in energy of both valence and conduction bands due to which CBB and VBT move upward and downward respectively in energy. The PDOS plot for the compound at $E_z$ = 5 V/nm is shown in Figure 3. When compared with the plot without $E_z$, it can be observed that there is decrease in Cd 5s, Tl 6s and Te 5p states in the valence band and increase in Cd 5s, Tl 6p and Te 5p states in the conduction band. However, the contributions of states for VBT are same and for CBB, Tl 6s states are more dominant. Also, along with increase and decrease of states in CBB and VBT respectively, the portion with higher density of states moves away from the Fermi level in both cases in presence of $E_z$. The obtained band gap with $E_z$ suggests that $CdTl_2Te_4$ will not be optically transparent as there will be some amount of free carriers created by absorption.

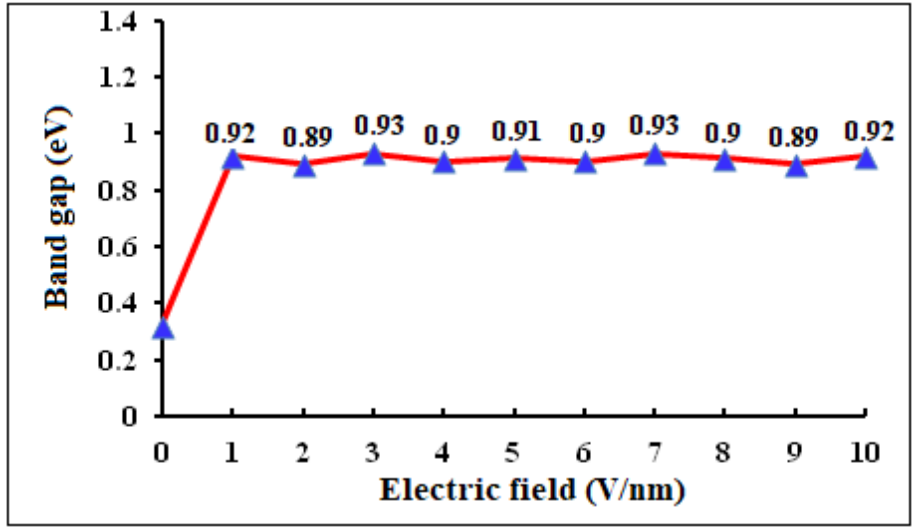

Figure 4: Plot showing variation of band gap ($\Delta_g$) with electric field ($E_z$). Energy gap at respective $E_z$ strength is mentioned

**Optical properties**

As mentioned in the computational methods, the optical properties of $CdTl_2Te_4$ have been calculated for polarized light along both in plane ($\mathbf{E}\perp c$) and out of the plane ($\mathbf{E}\|c$) directions of light and the properties which are calculated are real ($\varepsilon_1$) & imaginary ($\varepsilon_2$) parts of dielectric function, absorption coefficient ($\alpha$), reflectance ($R$), extinction coefficient ($k$) and refractive index ($n$) in the energy range 0 - 30eV. Firstly, these properties are deduced in absence of the external electric field for the unit cell of the compound. It is observed that the properties showed similarity for both the directions of polarization of light, from which it can be said that for this ternary chalcogenide, the optical properties are not anisotropic, that is does not depend on direction of polarized light. Further, to study the effect of externally applied field on these properties, $E_z$ of magnitudes 2.5, 5, 7.5 and 10 V/nm was employed in the structure. Now, it is observed that the applied $E_z$ could not effectively modulate the properties for $\mathbf{E}\perp c$ direction whereas there were significant shifts in the region of energies with highest peaks and properties in the case of $\mathbf{E}\|c$ direction. However, it is noticed that there is no change in the optical properties with increase in magnitude of electric field strength as it seems that the increase in $E_z$ cannot facilitate further interband transitions. As only out of the plane direction of polarized light tends to cause shifts in energies, the optical properties along that direction with and without the electric field are shown and discussed further.

The real and imaginary parts of the dielectric function is obtained from the expression $\varepsilon(\omega) = \varepsilon_1(\omega) + i\varepsilon_2(\omega)$, where $\varepsilon(\omega)$ is the complex dielectric function. From Figure 5, it can be seen that without $E_z$, the highest peak of $\varepsilon_1$ occurs at 1.46 eV which lies in the near infrared (IR) region (1.2 – 1.8 eV) with the maximum value 21.89. Also, in the range ~5-15 eV, it can be observed that due to plasmonic excitations, $\varepsilon_1$ attains negative values which shows that the material behaves as a metal in this energy range and also that most of the incident light is reflected in this range. When $E_z$ is applied, although a dip in the maximum $\varepsilon_1$ value is observed but there is a blue shift (~ 3eV) of energy with the highest peak which just passed the visible region (1.8 – 3.9 eV) and lies in the ultraviolet (UV) region (4.0 - 14 eV). Figure 5, which illustrates the energy range and peaks for $\varepsilon_2$, shows that when $E_z = 0$, the peak for maximum $\varepsilon_2$ value of 22.11 occurs at 1.64 eV, which is in near IR range. Applying $E_z$, it is seen that as similar to $\varepsilon_1$, there is a dip in the value but a large blue shift of ~4 eV is obtained and the region lies in UV range. Also with $E_z$, for lower energy range of 0 – 1.9 eV (majorly in IR range and some portion in visible range), the $\varepsilon_2$ value is almost zero, suggesting no absorbance in this region. For higher energy, both $\varepsilon_1$ and $\varepsilon_2$ values is seen to vanish which indicates that the compound does not interact with photons at higher frequency. The static dielectric constant (at zero energy) is 15.8 with no field and 4.7 in the presence of $E_z$.

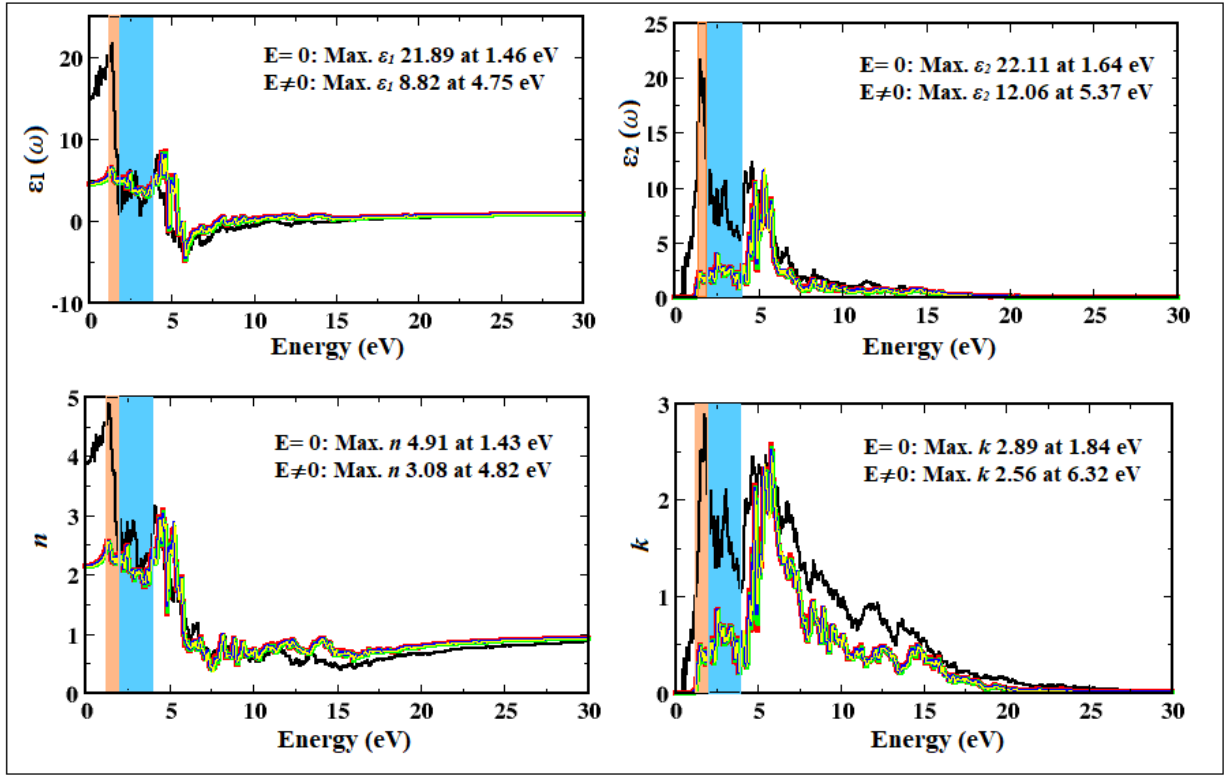

Figure 5: Optical spectra of real ($\varepsilon_1$) and imaginary ($\varepsilon_2$) parts of dielectric function, refractive index ($n$) and extinction coefficient ($k$). Black line denotes spectra when $E_z = 0$ and the coloured lines when $E_z \neq 0$ (*since all field strength show similar variation in properties, lines for particular $E_z$ values are indistinguishable*). The visible range is shown by the blue region and near IR range is shown by the orange region.

Figure 5 also shows the refractive index $n$ and extinction coefficient of $CdTl_2Te_4$ respectively. The static refractive index is 3.88 without electric field. The highest peak of $n$ is attained at 1.43 eV in the near IR regime with the maximum value 4.91. The value of $n$ drops to ~3 in the visible region and then with increase reduces heavily. In the presence of $E_z$, the static value of $n$ is reduced to 2.17 and having an energy shift, it attains a maximum peak at 4.82 eV in the UV region. In the visible region, its ranges between 2.21 to 2.54 and further saturates in the far UV range. The spectrum for extinction coefficient $k$ without $E_z$ shows that it attains the highest peak in the visible range at 1.84 eV, which tells that the penetration depth of the incident photons would be the least at this energy. There are some prominent peaks for $k$ in the UV region and with increase in energy it tends to zero. By applying external field, there is shift in the energy with the highest peak which occurs at 6.32 eV in the UV range. Also, the $k$ value is zero in the energy range 0 – 1.8 eV and the value also decreases rapidly in the visible regime in the presence of $E_z$. Similar to without $E_z$, with increase in energy, $k$ becomes almost zero.

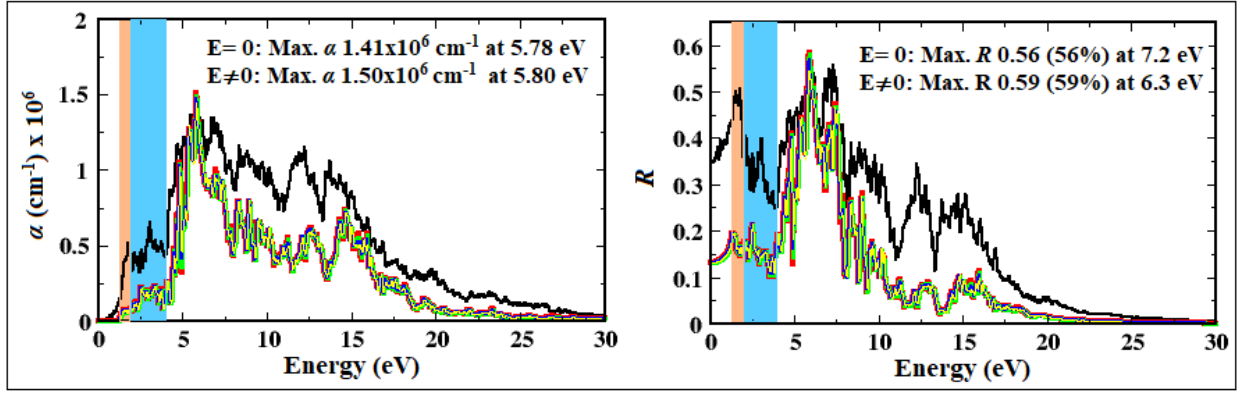

Figure 6: Optical spectra of absorption coefficient ($\alpha$) and reflectance ($R$)

Figure 6 shows the absorption coefficient ($\alpha$) of the compound and tells that there is strong absorption in the UV range in the absence of $E_z$ with the highest peak at 5.78 eV and $\alpha = 1.41 \times 10^6$ cm$^{-1}$, also suggesting there are low electron losses in this range. When $E_z$ is applied, there is no shift in energy is observed and also there is marginal increase in maximum value of $\alpha$, which is $1.50 \times 10^6$ cm$^{-1}$ at 5.80 eV. But one thing which must be noticed is $\alpha$ value becomes zero at the energy range 0 – 1.8 eV, which was observed for $\varepsilon_2$ earlier. Also the absorbance is quite low in the visible region, which shows the visible light can pass almost easily through this compound.
The reflectance ($R$) of the compound from Figure 6 shows that the highest peaks lie in both visible ($R = 0.5$) and UV regions ($R = 0.56$) with the highest peak at 7.2 eV when $E_z = 0$. Employing electric field, there is a red shift of ~1 eV and the $R$ marginally increased to 0.59. But, in the visible region, the reflectance is effectively reduced with $R$ values ranging around 0.15 and 0.22. So, the compound reflects only 15% - 22% incident light in the visible region.

The shifts in energy and the reduction in values of the calculated optical properties is mainly related to the increase in interband transition and subtle movement of the states as observed from PDOS plot (Figure 3) caused due to the electric field. The obtained optical properties with and without the electric field are interesting as it may find many possible applications. The high dielectric constant suggests that the material has high polarizability and one can also observe the contrasting behavior of dielectric function with regard to different ranges of electromagnetic spectrum. The value of $n > 3$ in the UV range, both with and without $E_z$ suggests that the material can be used as inner layer coating between UV absorbing layer and the substrate. Also, with $E_z$, the range of $n$ values in visible region is in good agreement to a previous finding that this can be used in solar cells as hole transport material [42]. The peaks of absorption spectra without $E_z$ in the UV range suggest that this compound can be used to protect sensitive devices from exposure to UV radiations. Also, with $E_z$ the low absorbance in the visible and near IR region, tells this material will not absorb much visible light and will pass through the material without attenuation. The reflectivity of this compound shows it can be used as selective reflectors in IR and UV regions. Also in presence of electric field, with very less reflectivity in visible range, it supports that the compound will absorb most of the light. So, the values of $n$, lower values of $\varepsilon_2$, $k$ and $\alpha$, very less reflectivity in the visible range in the presence of electric field suggest that $CdTl_2Te_4$ can be used efficiently in the solar cells as it can absorb most of the incident solar light.

**Conclusion**

As a summary, a $AB_2X_4$ type of ternary chalcogenide $CdTl_2Te_4$ has been studied and the electronic and optical properties of the compound were calculated using DFT implemented software SIESTA including modulation of the obtained properties via external electric field. The compound has a tetragonal crystal structure and taking the unit cell which is periodic in all directions optimization was done firstly. With the relaxed geometry, electronic and optical properties were calculated. Further, creating a vacuum in the unit cell along z direction, electric field along $c$ axis was applied. The obtained band structure showed that the compound is a narrow-gap semiconductor ($\Delta_g$ = 0.32 eV) and from the PDOS plot it is seen that the Te 5p and Tl 6p states play a major role for the energy gap. With $E_z$, the gap increases maximum upto 0.93 eV and remains almost same with all the fields applied. For optical properties without $E_z$, the maximum peaks of $R$ in UV & near IR range and $\alpha$ in UV range suggest it can be used as UV absorbers and selective reflectors. With $E_z$, the lower values of $R$, $\alpha$ and $\varepsilon_2$ in the visible region shows that it has a potential to be used to increase efficiency in solar cells. Also the higher refractive index obtained in UV region can find suitable usage. Hence, the compound is observed to possess various useful properties which can be tuned using external field. It is believed that these theoretical calculations when experimentally carried out will be of use in various fields like electronics, optoelectronics, etc.

**Conflict of interest**

There are no underlying conflicts of interest with this article.